\documentstyle[12pt,fleqn]{article}

\def\thebibliography#1{\section*{
References}\list
  {\arabic{enumi}.}{\settowidth\labelwidth{#1}\leftmargin\labelwidth
    \advance\leftmargin\labelsep
    \usecounter{enumi}}
    \def\newblock{\hskip .11em plus .33em minus .07em}
    \sloppy\clubpenalty4000\widowpenalty4000
    \sfcode`\.=1000\relax}

\let\Large=\large

\def\op#1{\mathop{\fam0 #1}\limits}

\newcommand{\Ker}{{\rm Ker\,}}
\newcommand{\im}{{\rm Im\, }}

\newcommand{\nm}[1]{\mid {#1}\mid}

\newcommand{\beq}{\begin{equation}}
\newcommand{\eeq}{\end{equation}}
\newcommand{\ben}{\begin{eqnarray}}
\newcommand{\een}{\end{eqnarray}}
\newcommand{\be}{\begin{eqnarray*}}
\newcommand{\ee}{\end{eqnarray*}}
\newcommand{\bea}{\begin{eqalph}}
\newcommand{\eea}{\end{eqalph}}

\newcommand{\cO}{\Omega}

\newcommand{\bR}{{\bf R}}

\newcommand{\bN}{{\bf N}}

\newcommand{\la}{\lambda}
\newcommand{\La}{\Lambda}
\newcommand{\f}{\phi}

\newcommand{\Om}{\Omega}
\newcommand{\m}{\mu}

\newcommand{\g}{\gamma}

\newcommand{\ve}{\varepsilon}
\newcommand{\th}{\theta}

\newcommand{\si}{\sigma}
\newcommand{\Si}{\Sigma}

\newcommand{\w}{\wedge}

\newcommand{\ol}{\overline}

\newcommand{\dr}{\partial}

\newcommand{\ar}{\op\longrightarrow}

\newcounter{eqalph}
\newcounter{equationa}
\newcounter{example}
\newcounter{remark}
\newcounter{theorem}
\newcounter{proposition}
\newcounter{lemma}
\newcounter{corollary}
\newcounter{definition}

\def\theremark{\arabic{remark}}

\def\thedefinition{\arabic{definition}}

\newenvironment{proof}{\noindent {\it Proof.}}{\hfill $\Box$
\medskip }

\newenvironment{rem}{\refstepcounter{remark} \medskip\noindent  REMARK
\theremark.}{ \medskip }
\newenvironment{theo}{\refstepcounter{definition} \medskip\noindent
THEOREM \thedefinition.\it}{\medskip }
\newenvironment{prop}{\refstepcounter{definition} \medskip\noindent
PROPOSITION \thedefinition.\it}{\medskip }
\newenvironment{lem}{\refstepcounter{definition} \medskip\noindent  LEMMA
\thedefinition.\it }{\medskip }

\newenvironment{eqalph}{\stepcounter{equation}
\setcounter{equationa}{\value{equation}}
\setcounter{equation}{0}

\begin{eqnarray}}{\end{eqnarray}\setcounter{equation}{\value{equationa}}}

\begin{document}
\hbox{}

{\parindent=0pt 

{ \Large \bf On the Obstruction to the Exactness of the Variational Bicomplex}
\bigskip

{\sc G.Giachetta$^\dagger$\footnote{giachetta@campus.unicam.it},
L.Mangiarotti$^\dagger$\footnote{mangiaro@camserv.unicam.it}  and
G.Sardanashvily$^\ddagger$}\footnote{sard@campus.unicam.it;
sard@grav.phys.msu.su} 

{ \small

{\it $^\dagger$ Department of Mathematics and Physics, University of Camerino,
62032 Camerino (MC), Italy

$^\ddagger$ Department of Theoretical Physics,
Physics Faculty, Moscow State University, 117234 Moscow, Russia}
\bigskip

{\bf Abstract.} We study the obstruction to the exactness of the variational
complex for a field theory on an affine bundle.  
\medskip

} }

\section{Introduction}

Let $Y\to X$ be a real smooth fibre bundle coordinated by
$(x^\la,y^i)$, and
$\cO^*_\infty$ the algebra of exterior forms on all finite order jet manifolds
of
$Y\to X$ modulo the pull-back identification. Put dim$X=n$. We have
the variational complex
\beq
0\to\bR\to \cO^0_\infty \ar^{d_H}\cO^{0,1}_\infty\ar^{d_H}\cdots  
\op\longrightarrow^{d_H} 
\cO^{0,n}_\infty  \op\longrightarrow^{\ve_1} E_1 
\op\longrightarrow^{\ve_2} 
E_2 \longrightarrow \cdots, \label{b317}
\eeq
where $\cO^{0,*}_\infty$ is a subalgebra of horizontal (semibasic) exterior
forms, $d_H$ is the horizontal exterior differential, $\ve_1$ is the
Euler--Lagrange map, $\ve_2$ is the Helmholtz--Sonin map, and so on 
\cite{ander,bau,dedt,dick,book,tak2,tul}. 
This complex provides the algebraic approach
to the calculus of variations in field theory. 
The well-known theorem states the local exactness of this complex as follows.

\begin{theo} \label{lmp4} 
If a fibre bundle
$Y$ is
$\bR^{n+m}\to \bR^n$, the variational complex (\ref{b317}) except the first
term is exact, i.e.
\be
(i)\,\Ker d_H= \im d_H, \qquad (ii)\,\Ker\ve_1=\im d_H, \qquad (iii)\,\Ker
\ve_{k+1}=\im\ve_k.
\ee
\end{theo}

\begin{rem}
As in the De Rham complex on a connected manifold, one puts $H^0(d_H)=\bR$ 
because, if $d_Hf=0$, $f\in\cO^0_\infty$, then $f=$const. 
From now on, by
$H^*$ we will mean the cohomology groups of order $k>0$.
\end{rem}

Restricted to $\cO^{0,*}_\infty$, the variational complex (\ref{b317}) yields
the so called horizontal complex 
\beq
0\to\bR\to \cO^0_\infty \ar^{d_H}\cO^{0,1}_\infty\ar^{d_H}\cdots  
\op\longrightarrow^{d_H} 
\cO^{0,n}_\infty  \op\longrightarrow^{d_H} 0. \label{+481}
\eeq
Then a corollary of Theorem \ref{lmp4} is the algebraic Poincar\'e lemma on the
local exactness of the complex (\ref{+481}). 

\begin{prop} If a fibre bundle
$Y$ is $\bR^{n+m}\to \bR^n$, the horizontal complex has the cohomology groups
\beq
 H^{k<n}(d_H)=0, \qquad
H^n(d_H)=\ve_1(\Om^{0,n}_\infty). \label{lmp5}
\eeq
\end{prop}

This work is devoted to the obstruction to the
(global) exactness of the complexes (\ref{b317}) and (\ref{+481}). 

It should be emphasized that we consider the variational bicomplex in the
class of smooth exterior forms $\cO^*_\infty$ on the infinite-order jet space
$J^\infty Y$ which are the pull-back of exterior forms on finite-order
jets. This is not the case of \cite{ander0,ander}. In these works 
smooth forms on $J^\infty Y$ which are locally the pull-back of
forms on finite-order jet manifolds are considered. The key point is that 
$J^\infty Y$ admits a partition of unity with respect to this class of
smooth functions. Then, applying the well-known Mayer-Vietoris
sequence
\cite{bott}, one can show that contact $d_H$-closed forms are
necessarily
$d_H$-exact. This may not be true for the class of exterior forms
$\cO^*_\infty$ which usually are utilized in field theory.

We show
that, if a fibre bundle $Y\to X$ admits a global section (e.g., if it is an
affine bundle), there is a monomorphism of the De Rham homology groups
$H^*(X)$ of the base manifold
$X$ to the cohomology groups $H^*_{\rm var}$  
of the variational complex (\ref{b317}) and to those $H^*(d_H)$ of the
horizontal complex (\ref{+481}). When $Y\to X$ is a vector bundle, we find the
obstruction to the exactness  of the variational complex (\ref{b317}) at 
$E_k$. We show that, in this case, the cohomology group of the variational
complex (\ref{b317}) at $\cO^{0,n}_\infty$ is equal to the
cohomology group $H^n(X)$ of the base manifold. It states the following
theorem (see also \cite{kru98}).

\begin{theo} \label{lmp13} If $Y\to X$ is an affine bundle, an
$r$-order Lagrangian
$L\in
\cO^{0,n}_r$ is variationally trivial iff it takes the form $L=d_H\si$ where
$\si\in
\cO^{0,n-1}_{r-1}$ is a horizontal $(n-1)$-form on the jet manifold
$J^{r-1}Y$. 
\end{theo}

Recall that the familiar result states that $L$ is locally $d_H$-exact.

\section{Preliminaries}

Unless otherwise stated,  manifolds throughout are assumed to be real, smooth,
finite-dimensional, Hausdorff, paracompact and connected. 

Let $J^rY$ be the $r$-order jet manifold  of
sections of a fibre bundle $Y\to X$. It is endowed with the adapted coordinates
$(x^\la, y^i_\La)$, $0\leq\nm\La \leq r$, 
where $\La=(\la_k...\la_1)$, $|\La|=k$,
is a symmetric multi-index. By $\La+\Si$ we will denote the symmetric
multi-index
$(\la_k\cdots\la_1\si_m\cdots\si_1)$.
The transition functions of these coordinates read
\beq
{y'}^i_{\la+\La}=\frac{\dr x^\m}{\dr x'^\la}d_\m y'^i_\La, \label{55.21}
\eeq
where by $d_\la$ are meant 
the total derivatives 
\be
d_\la = \dr_\la + \op\sum_{|\La|=0} y^i_{\La+\la}\dr_i^\La.
\ee
We will use the notation 
\be
\dr_\La=\dr_{\la_k}\circ\cdots\circ\dr_{\la_1}, \qquad
 d_\La=d_{\la_k}\circ\cdots\circ d_{\la_1}, \qquad
\La=(\la_k...\la_1).
\ee

There is the inverse system
\beq
X\op\longleftarrow^\pi Y\op\longleftarrow^{\pi^1_0}\cdots \longleftarrow
J^{r-1}Y \op\longleftarrow^{\pi^r_{r-1}} J^rY\longleftarrow\cdots 
\label{5.10}
\eeq
of fibrations (surjective submersions) of finite order jet manifolds. This
inverse system has  a projective limit 
$J^\infty Y$. It is a paracompact Fr\'echet manifold which
admits a smooth partition of unity
\cite{abb,bau,tak1,tak2}. Differential objects on $J^\infty Y$ are
introduced as operations on the $\bR$-ring $C^\infty(J^\infty Y$) of smooth
locally cylindrical functions. A real function
$f$ on $J^\infty Y$ is said to be a smooth (locally cylindrical)
function if it is locally a pull-back
of a smooth function on some finite-order jet manifold $J^kY$ with
respect to the surjection $\pi^\infty_k: J^\infty Y\to J^kY$. 
Vector fields on  $J^\infty Y$ are defined as derivations of this ring.
They make up the left locally free $C^\infty(J^\infty Y)$-module
Der$\,(C^\infty(J^\infty Y))$. The $C^\infty(J^\infty Y)$-dual of 
Der$(C^\infty(J^\infty Y))$ is the 
$C^\infty(J^\infty Y)$-module of exterior 1-forms on
$J^\infty Y$. In field theory, 
one usually considers the direct limit $\cO^*_\infty$ in the category of
$\bR$-modules of the direct system 
\beq
\cO^*(X)\op\longrightarrow^{\pi^*} \cO^*(Y) 
\op\longrightarrow^{\pi^1_0{}^*} \cO_1^*
\op\longrightarrow^{\pi^2_1{}^*} \cdots \op\longrightarrow^{\pi^r_{r-1}{}^*}
 \cO_r^* \longrightarrow\cdots. \label{5.7}
\eeq
of the $\bR$-modules $\cO^*_k$ of exterior forms on finite-order jet manifolds
$J^kY$. It consists of pull-backs of exterior forms on finite-order jet
manifolds.  

The $\bR$-module $\cO_\infty^*$ possesses the structure of a
filtered module given by the $\cO^0_k$-submodules $\cO_k^*$ of
$\cO^*_\infty$ \cite{vinb}.
An endomorphism $\g$ of
$\cO_\infty^*$ is called a filtered morphism if there exists $i\in \bN$
such that the restriction of 
$\g$ to $\cO_k^*$ is the homomorphism of $\cO_k^*$ into  $\cO_{k+i}^*$
over the injection $\cO^0_k\hookrightarrow\cO^0_{k+i}$ for all $k$. 

\begin{theo} \label{+468}  \cite{massey}. 
Every direct system of endomorphisms $\{\g_k\}$ of $\cO_k^*$
such that 
\be
\pi^j_k{}^*\circ\g_k = \g_j\circ\pi^j_k{}^* 
\ee
for all $j>k$ has the  direct limit $\g_\infty$ in filtered endomorphisms of
$\cO_\infty^*$.  If all $\g_k$ are
monomorphisms [epimorphisms], then $\g_\infty$ is also  a monomorphism
[epimorphism]. 
\end{theo}

By virtue of this theorem, the standard algebraic operations on exterior
forms on finite-order jet manifolds have the direct limits on
$\cO_\infty^*$, and make 
$\cO_\infty^*$ an exterior algebra.

Since $\cO^*_\infty$ consists of pull-back forms, its elements 
can be written in the familiar coordinate form.
In particular, the basic 1-forms $dx^\la$ and the contact 1-forms
$\th^i_\La=dy^i_\La -y^i_{\la+\La}dx^\la$ provide 
the local generators of $\cO_\infty^*$, and define
the canonical splitting of the space of $m$-forms
\beq
\cO^m_\infty =\cO^{0,m}_\infty\oplus
\cO^{1,m-1}_\infty\oplus\ldots\oplus\cO^{m,0}_\infty,
\label{5.20}
\eeq
in the spaces $\cO^{k,m-k}$ of $k$-contact forms.
We have the corresponding $k$-contact projections
\be
h_k: \cO^m_\infty\to \cO^{k,m-k}_\infty, \qquad 1\leq k\leq m, 
\ee
and the horizontal projection
\be
h_0: \cO^m_\infty\to \cO^{0,m}_\infty.
\ee

Accordingly, the
exterior differential on $\cO_\infty^*$ is
decomposed into the sum $d=d_H+d_V$
of horizontal 
 and vertical differentials
\be
&& d_H:\cO_\infty^{k,s}\to \cO_\infty^{k,s+1}, \qquad d_H(\f)= dx^\la\w
d_\la(\f), \qquad \f\in\cO^*_\infty,\\ 
&& d_V:\cO_\infty^{k,s}\to \cO_\infty^{k+1,s}, \qquad
d_V(\f)=\th^i_\La \w \dr_\La^i\f.
\ee
They obey the nilpotency rule
\beq
d_H\circ d_H=0, \qquad d_V\circ d_V=0, \qquad d_V\circ d_H
+d_H\circ d_V=0 \label{lmp50}
\eeq
and the relations
\ben
&& h_0d=d_Hh_0, \label{hn1}\\
&& h_kdh_k=d_Hh_k, \label{hn1'}
\een

\section{Cohomology of the infinite-order De Rham complex}

The exterior algebra $\cO^*_\infty$ provides the infinite-order De Rham
complex  
 \beq
0\longrightarrow \bR\longrightarrow
\cO^0_\infty\op\longrightarrow^d\cO^1_\infty\op\longrightarrow^d
\cdots\,.
\label{5.13}
 \eeq

\begin{prop} \label{lmp1}
The cohomology $H^*(\cO^*_\infty)$ of the infinite-order De Rham complex
(\ref{5.13}) coincides
with the De Rham cohomology $H^*(Y)$ of the fibre bundle $Y$.
\end{prop}

The result is a corollary of the following two assertions. 

\begin{prop}\label{ch511} \cite{massey}. The operation of taking
cohomology groups of a cochain complex commutes with the passage to
the direct limit.
\end{prop}

It is a corollary of Theorem \ref{+468}.

\begin{prop} \label{lmp2} \cite{bau}.
The De Rham cohomology $H^*(J^kY)$ of any finite-order jet manifold $J^kY$ is
equal to that of the fibre bundle $Y$.
\end{prop}

It is based on the fact that jet bundles $J^kY\to J^{k-1}Y$ are
affine, while the De Rham cohomology of an affine bundle is equal to that of
its base.

Recall that, given a fibre bundle $\pi:Y\to X$, there is a homomorphism
\beq
\pi^*: H^*(X)\to H^*(Y) \label{lmp21}
\eeq
of the De Rham cohomology groups of its base $X$ to those of $Y$. 

\begin{lem} If a fibre bundle $Y\to X$ admits a global
section $s$, there is the sequence of homomorphisms of cohomology groups
\beq
H^*(X)\ar^{\pi^*} H^*(Y) \ar^{s^*} H^*(X), \label{lmp22}
\eeq
where $s^*$ is an epimorphism and, consequently, $\pi^*$ (\ref{lmp21}) is a
monomorphism.
\end{lem}

If $Y\to X$ admits a global section, we have the corresponding  monomorphism 
\beq
\pi^*: H^*(X)\hookrightarrow H^*(\cO^*_\infty) \label{lmp20}
\eeq
of the De Rham cohomology groups of the base $X$
to the cohomology groups of the infinite-order De Rham complex (\ref{5.13}). 

If $Y\to X$ is
an affine bundle, the monomorphism (\ref{lmp20}) is an isomorphism 
\beq
H^*(\cO_\infty^*)=H^*(X). \label{hn2}
\eeq
In this case, any closed form $\f\in\cO^*_\infty$
is decomposed into the sum 
\beq
\f=\varphi +d\xi \label{lmp9}
\eeq 
where $\varphi\in \cO^*(X)$ is a closed form on $X$.

\section{Cohomology of the horizontal complex}

Due to the nilpotency rule (\ref{lmp50}), the vertical and horizontal
differentials $d_V$ and $d_H$ define the bicomplex
\beq
\begin{array}{rcrlcrlcr}
& & & \put(0,-10){\vector(0,1){20}} & & & \put(0,-10){\vector(0,1){20}} & & \\
\cdots &\ar & & \cO^{k+1,m}_\infty & \ar^{d_H} & & \cO^{k+1,m+1}_\infty
&\ar &\cdots \\ 
& & _{d_V} & \put(0,-10){\vector(0,1){20}} & & _{d_V} &
\put(0,-10){\vector(0,1){20}} & & \\
\cdots &\ar & & \cO^{k,m}_\infty & \ar^{d_H} & &
\cO^{k,m+1}_\infty & \ar & \cdots\\ 
& & & \put(0,-10){\vector(0,1){20}} & &
 & \put(0,-10){\vector(0,1){20}} & & 
\end{array}
\label{7}
\eeq
The rows and columns of these bicomplex are horizontal and vertical
complexes. Let us start from the vertical ones.

\begin{prop} \label{lmp53} 
If $Y\to X$ is a vector bundle, then any $d_V$-closed form $\f$ is the sum 
$\f=d_V\si +\varphi$ of a $d_V$-exact form and an exterior form on $X$,
i.e., vertical complexes are exact.
\end{prop}

\begin{proof}
Local exactness of a vertical complex on a coordinate chart
$(x^\la,y^i)$ follows from a version of the Poincar\'e lemma with
parameters. We have the the corresponding homotopy operator
\beq
\si=\int^1_0 t^k [ \ol y\rfloor\f(x^\la, ty^i_\la)]dt, \qquad \f\in \cO^{k,*},
\label{lmp51}
\eeq
where $\ol y=y^i_\La\dr_i^\La$. Since $Y\to X$ is a vector bundle, it is
readily observed that the homotopy operator is globally defined, and so is
the form $\si$.
\end{proof}

Turn now to the rows of the diagram (\ref{7}). We will start from the
horizontal complex (\ref{+481}). Due to the relation (\ref{hn1}), the
horizontal projection
$h_0$ defines a homomorphism of the infinite-order De Rham complex (\ref{5.13})
to the horizontal complex (\ref{+481}) which sends closed and exact forms to
$d_H$-closed and $d_H$-exact horizontal forms, respectively. Accordingly, it
yields the homomorphism of the cohomology groups of these complexes
\beq
h_0^*: H^*(\cO^*_\infty)\to H^*(d_H), \label{hn3}
\eeq
which is neither monomorphism nor epimorphism in general. By virtue of
Proposition \ref{lmp1}, the homomorphism (\ref{hn3}) is the
homomorphism of the cohomology groups
\beq
h_0^*: H^*(Y)\to H^*(d_H). \label{hn4}
\eeq
Then we also have the homomorphism
\beq
h_0^*\circ \pi^*:H^*(X)\to H^*(d_H). \label{hn5}
\eeq

\begin{prop} \label{lmp6}  
If a fibre bundle $Y\to X$ admits a global section $s$,
the homomorphism (\ref{hn5}) is a monomorphism.
\end{prop}

\begin{proof}
Bearing in mind the monomorphism (\ref{lmp20}), let $\f\in\cO^*_\infty$ be a
closed form whose cohomology class belongs to $H^*(X)\subset
H^*(\cO^*_\infty)$.  We will show that, if $h_0\f$ is $d_H$-exact, $\f$ is
exact. The form $\f$ is 
represented by the sum (\ref{lmp9}), and we have
\be
h_0\f=\varphi + d_H(h_0\xi).
\ee
 Let $\varphi=d_H\si$ where
$\si$ is an exterior form on some finite-order jet manifold $J^kY$. Given
a global section $s$ of the fibre bundle
$Y\to X$, let $J^{k+1}s$ be its $(k+1)$-order jet prolongation to a section
of the jet bundle $J^{k+1}Y\to X$. Then it
is readily observed that
\be
\varphi=(J^{k+1}s)^*(d_H\si)= (J^{k+1}s)^*(d\si)=d ((J^{k+1}s)^*\si). 
\ee
\end{proof}

Since $h_0$ is a homomorphism of the infinite-order De Rham complex
(\ref{5.13}) to the horizontal complex (\ref{+481}), we have the simple exact
sequence of the cochain complexes
\beq
\begin{array}{rcrlcrlcccrlclcc}
& & & 0 & & & 0 & & & &  & 0 & & & &  \\
& & & \put(0,10){\vector(0,-1){20}} & & & \put(0,10){\vector(0,-1){20}} & &
& & & \put(0,10){\vector(0,-1){20}}& & & &  \\
\cdots &\ar^d & & \Ker h_0 & \ar^d & & \Ker h_0 &\ar^d
&\cdots &\ar^d & & \Ker h_0 & \ar^d &\Ker h_0 & &\\ 
& & _{\rm in} & \put(0,10){\vector(0,-1){20}} & & _{\rm in} &
\put(0,10){\vector(0,-1){20}} & & & & _{\rm in} & 
\put(0,10){\vector(0,-1){20}} &  & || & &\\
\cdots &\ar^d & & \cO^{k-1}_\infty & \ar^d & &
\cO^k_\infty & \ar^d &  \cdots & \ar^d & & \cO^n_\infty & \ar^d &
\cO^{n+1}_\infty &\ar& \cdots\\ 
& &_{h_0} & \put(0,10){\vector(0,-1){20}} & &
_{h_0} & \put(0,10){\vector(0,-1){20}} & & & & _{h_0} &
\put(0,10){\vector(0,-1){20}} & & \put(0,10){\vector(0,-1){20}} & &\\
\cdots & \ar^{d_H} & & \cO^{0,k-1}_\infty & \ar^{d_H} & &
\cO^{0,k}_\infty & \ar^{d_H} &  \cdots & \ar^{d_H} & & \cO^{0,n}_\infty &
\ar & 0 & &\\ 
& & & \put(0,10){\vector(0,-1){20}} & & &
\put(0,10){\vector(0,-1){20}} & & & & &\put(0,10){\vector(0,-1){20}} & & &
& \\ 
& & & 0 & & & 0 & & & & & 0 & & & &
\end{array}
\label{6'}
\eeq
Its first row is a subcomplex of the infinite-order De Rham complex
(\ref{5.13}). There is the corresponding exact sequence of cohomology groups 
\beq
\cdots \ar H^k(\Ker h_0) \ar^{{\rm in}^*} H^k(\cO^*_\infty)
\ar^{h_0^*} H^k(d_H) \ar^{d^*}  H^{k+1}(\Ker h_0) \ar \cdots\,. \label{lmp25}
\eeq
We can say something more on this exact sequence when $Y\to X$ is an affine
bundle.

\begin{prop}
If $Y\to X$ is an affine bundle, the exact sequence of cohomology groups
(\ref{lmp25}) takes the form
\ben
&& \cdots \ar H^{k-1}(d_H)/H^{k-1}(X) \ar^{{\rm in}^*} H^k(X)
\ar^{h_0^*} H^k(d_H) \ar^{d^*}  H^k(d_H)/H^k(X) \ar  \label{lmp26}\\
&& \qquad \cdots \ar^{{\rm in}^*} H^n(X)
\ar^{h_0^*} H^n(d_H) \ar^{d^*}  H^n(d_H)/H^n(X) \ar 0, \nonumber
\een
where ${\rm in}^*$ is a homomorphism to $0\in H^*(X)$ and $h_0^*$ is a
monomorphism.
\end{prop}

\begin{proof}
Firstly, let us describe the cohomology group $H^k(\Ker h_0)$ of the first
row in the diagram (\ref{6'}). In this complex, the cocycles  are closed
forms $\f\in \Ker h_0$, while the coboundaries are exact forms
$d\xi$ where $\xi\in \Ker h_0$. Because of the isomorphism (\ref{hn2}),
such a $k$-cocycle is a coboundary in the infinite-order De Rham complex and is
decomposed into the sum 
\beq
\f=d\si +d\xi, \label{lmp27}
\eeq
where $\si \in \cO^{0,k-1}_\infty$ is a horizontal form such that $h_0
d\si=0$. It follows that $d_H\si=0$, i.e., $\si$ is a $d_H$-closed form. At
the same time, if
$\si$ is a $d_H$-exact form $\si=d_H\psi$, then 
\be
d\si=d d_H\psi=-d d_V\psi
\ee
is a coboundary in $\Ker h_0$. Therefore, we have an epimorphism of
cohomology groups
\be
d^*: H^{k-1}(d_H) \to H^k(\Ker h_0).
\ee
Obviously, its kernel $\Ker d^*$ contains the cohomology group
$H^{k-1}(X)\subset H^{k-1}(d_H)$. We show that $\Ker d^*=H^{k-1}(X)$.
Let $\si\in \Ker d^*$, i.e.,
\be
d_H\si=0, \qquad d\si=\psi, \qquad \psi\in \Ker h_0.
\ee
Then we have $d(\si-\psi)=0$, and
\be
\si =\varphi + d\xi + \psi, \qquad \varphi \in \cO^*(X),
\ee 
in accordance with the decomposition  (\ref{lmp9}). Moreover, being
horizontal, 
\be
\si=\varphi + h_0d\xi=\varphi + d_Hh_0\xi,
\ee
i.e., it belongs to the $d_H$-cohomology class of the form $\varphi$ on $X$.
At last, since all cocycles in $\Ker h_0$ are coboundaries in $\cO^*_\infty$,
the morphism ${\rm in}^*$ sends the cohomology group $H^k(\Ker h_0)$ to zero.
\end{proof}

Turn now to cohomology of the $k$-contact horizontal complex
\beq
\cdots\ar^{d_H} \cO_\infty^{k,m-1}\ar^{d_H} \cO_\infty^{k,m}\ar^{d_H}\cdots.
\label{lmp55}
\eeq
It is readily observed that, because of the relation (\ref{hn1}), the contact
projection $h_k$ fails to be a homomorphism of the infinite-order De Rham
complex (\ref{5.13}) to the horizontal complex (\ref{lmp55}). Accordingly, we
have no homomorphism of the corresponding cohomology groups. At the same
time, due to the nilpotency rule (\ref{lmp50}), the horizontal differential
yields a homomorphism of the $k$-contact horizontal complex (\ref{lmp55})
to the $(k+1)$-contact one, together with the corresponding homomorphism
of their cohomology groups
\be
d_V^*: H^*(k,d_H)\to H^*(k+1,d_H).
\ee
Moreover, we have the complex of the cohomology groups
\beq
0\ar H^*(X)\ar^{h_0^*}H^*(d_H) \ar^{d_V^*}\cdots \ar^{d_V^*}
H^*(k,d_H)\ar^{d_V^*}\cdots\,. \label{lmp56}
\eeq
Its cohomology is called the second cohomology \cite{mcl}. 

Similarly, the horizontal differential $d_H$ defines homomorphisms of the
vertical complexes.

Let us show that, if the fibre bundle $Y\to X$ admits a global section, the
complex (\ref{lmp56}) is exact at the second term, i.e. the kernel of $d_V$
in $H^*(d_H)$ coincides with the De Rham cohomology $H^*(X)$.

\begin{lem} \label{lmp63} 
If $\f\in\Ker d_Hd_V$, then 
\beq
\f=\si +\xi +\varphi, \label{lmp57}  
\eeq
where $\si$ is a $d_H$-closed form, $\xi$ is a $d_V$-closed form and
$\varphi\in \cO^*(X)$ (see \cite{tul} for the local case).
\end{lem}

\begin{prop} \label{lmp60} 
If the fibre bundle $Y\to X$ admits a global section, the
complex (\ref{lmp56}) is exact at the second term. 
\end{prop}

\begin{proof}
Let $\f\in\cO^{0,*}_\infty$ be a $d_H$-closed form, i.e., $d_H\f=0$. It
belongs to $\Ker d_V^*$ if $d_V\f=d_H\ve$. In accordance with the relation
(\ref{lmp57}), it follows that $\f=d_H\varepsilon +\varphi$, i.e. belongs to
$H^*(X)\subset H^*(d_H)$.
\end{proof}

\section{Cohomology of the variational complex}

Obviously, the cohomology groups $H^{<n}_{\rm var}$ of the variational complex
(\ref{b317}) coincide with those the horizontal complex
(\ref{+481}). To say something on other cohomology groups $H^{\geq n}_{\rm
var}$, let us consider the simple exact sequence of cochain complexes
\beq
\begin{array}{rcrlcrlcrlcrlcl}
& & & 0 & & & 0 & & &0 & & &0 & &\\
& & & \put(0,10){\vector(0,-1){20}} & & & \put(0,10){\vector(0,-1){20}}
 & & & \put(0,10){\vector(0,-1){20}} & & & \put(0,10){\vector(0,-1){20}} & &\\
\cdots &\longrightarrow & & \Ker h_0 & \op\longrightarrow^d & &
\Ker h_0 & \op\longrightarrow^d &  & \Ker e_1&
\op\longrightarrow^d & &
\Ker e_2 & \longrightarrow & \cdots \\
& & & \put(0,10){\vector(0,-1){20}} & & & \put(0,10){\vector(0,-1){20}}
 & & & \put(0,10){\vector(0,-1){20}} & & & \put(0,10){\vector(0,-1){20}} & &\\
\cdots &\longrightarrow & & \cO^{n-1}_\infty & \op\longrightarrow^d & &
\cO^n_\infty & \op\longrightarrow^d & & \cO^{n+1}_\infty &
\op\longrightarrow^d & &
\cO^{n+2}_\infty & \longrightarrow & \cdots \\
& &_{h_0} & \put(0,10){\vector(0,-1){20}} & & _{h_0}
& \put(0,10){\vector(0,-1){20}}
 & & _{e_1}& \put(0,10){\vector(0,-1){20}} & & _{e_2}
& \put(0,10){\vector(0,-1){20}} & &\\
\cdots &\longrightarrow & & \cO^{0,n-1}_\infty & \op\longrightarrow^{d_H} & &
\cO^{0,n}_\infty & \op\longrightarrow^{\ve_1} & & E_1 &
\op\longrightarrow^{\ve_2} & &
E_2 & \longrightarrow & \cdots \\
& & & \put(0,10){\vector(0,-1){20}} & & & \put(0,10){\vector(0,-1){20}}
 & & & \put(0,10){\vector(0,-1){20}} & & & \put(0,10){\vector(0,-1){20}} & &\\
& & & 0 & & & 0 & & &0 & & &0 & &
\end{array}
\label{6}
\eeq
where $e_k=\tau_k\circ h_k$ and $\tau_k$ is the projection map providing the
decomposition
\be
\cO^{k,n}_\infty = \tau_k(\cO^{k,n}_\infty) + d_H(\cO^{k,n-1}_\infty).
\ee
We have the corresponding decomposition 
\beq
\cO^{n+k}_\infty=\Ker e_k\oplus E_k,\label{lmp40}
\eeq
where
\beq
\Ker e_k= d_H(\cO^{k,n-1}_\infty)\oplus(\op\oplus_{m>0} \cO^{k+m,n-m}).
\label{lmp62}
\eeq
It is readily observed that $\ve_k=\tau_k\circ d$ on $\cO^{k-1,n}_\infty$.
The diagram (\ref{6}) is derived from the variational bicomplex (see, e.g.,
\cite{book,tul}). Its first row is a subcomplexes of the infinite-order De Rham
complex (\ref{5.13}). 

Since the diagram (\ref{6}) coincides with the diagram
(\ref{6'}) at exterior forms of degree $\leq n$, the corresponding exact
sequence of cohomology groups for the diagram (\ref{6}) differs from the
exact sequence (\ref{lmp25}) starting from the terms after 
$H^n(\cO^n_\infty)$. If $Y\to X$ is an affine bundle, then
$H^{>n}(\cO^*_\infty)=0$ and the exact sequence at these terms breaks into the
short sequences
\ben
&& 0\ar H^n(X)\ar^{h_0^*} H^n_{\rm var} \ar^{d^*} H(\Ker e_1)\ar^{{\rm in}^*}
0, \label{lmp30}\\
&& 0\ar H^{n+k}_{\rm var}= H(e_{k+1})\ar 0, \qquad k> 0. \label{lmp31}
\een
Using these exact sequences and the definition (\ref{lmp62}) of
$\Ker e_k$, one can find the cohomology groups $H^{\geq n}_{\rm var}$ of the
variational complex (\ref{b317}). Recall that cocycles in $\Ker e_k$ are
closed forms
$\phi\in
\Ker e_k$, while the coboundaries are exact forms $d\xi$ where $\xi\in \Ker
e_{k-1}$. 

\begin{lem} \label{lmp61} 
The cohomology group $H(e_1)$ is trivial.
\end{lem}

\begin{proof}
The obstruction for the cohomology group $H(e_1)$ to be trivial are the
elements of the overlap
$d\cO^{0,n}_\infty\cap d_H\cO^{k,n-1}_\infty$, i.e., the forms $\f\in
\cO^{0,n}_\infty$ such that $d_V\f=d_H\ve$,
i.e., $\f\in \Ker d_Hd_V$. By virtue of Lemma \ref{lmp63}, such a form is
given by the sum (\ref{lmp57}) which reads
\be
\f=d_H\psi +\varphi.
\ee
Then, $d\f=-dd_V\psi$ is an exact form.
\end{proof}

It follows that the short exact sequence (\ref{lmp30}) leads to the
isomorphism 
\beq
 H^n(X)= H^n_{\rm var}. \label{lmp64}
\eeq
In other words, the cohomology group of variationally trivial Lagrangians on
an affine bundle modulo $d_H$-exact forms coincides with the cohomology group
$H^n(X)$ of the base manifold $X$. This states Theorem \ref{lmp13}.
The isomorphism (\ref{lmp64}) also leads to the isomorphism 
\be
H^n(d_H)/H^n(X)=\ve_1(\cO^{0,n}_\infty), 
\ee
where $H^n(d_H)$ is the $n$th-cohomology group of the horizontal complex
(\ref{+481}). It generalizes the local equality (\ref{lmp5}).

One can extend Lemma \ref{lmp61} to higher cohomology groups $H(e_k)$ as
follows.

\begin{prop}
The obstruction for the cohomology group $H(e_{k+1})$ and,
consequently, the cohomology group $H^{n+k}_{\rm var}$ to be trivial is the
second cohomology group of the complex (\ref{lmp56}) at the term
$H^k(d_H)$.
\end{prop}

The result follows from the decomposition (\ref{lmp40}).


\begin{thebibliography}{ederf}

\bibitem{abb} M.Abbati and A.Mani\`a, On differential structure for
projective limits of manifolds, {\it J. Geom. Phys.}, {\bf 29}, 35-63 (1999).

\bibitem{ander0} I.Anderson, Introduction to the variational bicomplex, {\it
Contemp. Math.}, {\bf 132}, 51-74 (1992).

\bibitem{ander} I.Anderson, {\it The Variational Bicomplex}, Academic Press,
Boston, 1994.

\bibitem{bau} M.Bauderon, Differential geometry and Lagrangian formalism in
the calculus of variations, in {\it Differential Geometry, Calculus of
Variations, and their Applications}, Lecture Notes in Pure and Applied
Mathematics, {\bf 100} (Marcel Dekker, Inc., N.Y., 1985), p. 67-82.

\bibitem{bott} R.Bott and L.Tu, {\it Differential Forms in Algebraic
Topology}, Springer-Verlag, Berlin, 1982.

\bibitem{dedt} P.Dedecker and W.Tulczyjew, Spectral sequences and the
inverse problem of the calculus of variations, in {\it Differential
Geometric Methods in Mathematical Physics}, Lect. Notes in Mathematics,
{\bf 836}, Springer-Verlag, Berlin, 1980, pp. 498-503.

\bibitem{dick} L.Dickey, On exactness of the variational bicomplex, {\it
Contem. Math.}, {\bf 132}, 307-315 (1992).

\bibitem{book} G.Giachetta, L.Mangiarotti and G.Sardanashvily, {\it New
Lagrangian and Hamiltonian Methods in Field Theory}, World Scientific,
Singapore, 1997.

\bibitem{vinb} I.Krasil'shchik, V.Lychagin and A.Vinogradov, {\it Geometry of
Jet Spaces and Nonlinear Partial Differential Equations}, Gordon and Breach,
Glasgow, 1985. 

\bibitem{kru98} D.Krupka, J.Musilova, Trivial Lagrangians in field theory,
{\it Diff. Geom. Appl.}, {\bf 9}, 293-305 (1998); Erratum, {\bf 10}, 303
(1999).

\bibitem{mcl} S.Mac Lane, {\it Homology} (Springer-Verlag, Berlin, 1967).

\bibitem{massey} W.Massey, {\it Homology and Cohomology Theory}, Marcel Dekker,
Inc., N.Y., 1978.

\bibitem{tak1} F.Takens, Symmetries, conservation laws and variational
principles, in {\it Geometry and Topology}, Lect. Notes in Mathematics, {\bf
597}, Springer-Verlag, Berlin, 1977, pp. 581-604.

\bibitem{tak2} F.Takens, A global version of the inverse problem of
the calculus of variations, {\it J. Diff. Geom.}, {\bf 14}, 543-562 (1979).

\bibitem{tul} W.Tulczyjew, The Euler--Lagrange resolution, 
in {\it Differential
Geometric Methods in Mathematical Physics}, Lect. Notes in Mathematics,
{\bf 836} (Springer-Verlag, Berlin, 1980), pp. 22-48.

\end{thebibliography}
\end{document}